\documentclass[preprint,12pt]{elsarticle}
\pdfoutput=1
 \usepackage{epsfig}

\usepackage{amsmath,amssymb}
\usepackage{float}
\usepackage{subfigure}
\begin{document}

\begin{frontmatter}

%% Title, authors and addresses

%% use the tnoteref command within \title for footnotes;
%% use the tnotetext command for theassociated footnote;
%% use the fnref command within \author or \address for footnotes;
%% use the fntext command for theassociated footnote;
%% use the corref command within \author for corresponding author footnotes;
%% use the cortext command for theassociated footnote;
%% use the ead command for the email address,
%% and the form \ead[url] for the home page:
%% \title{Title\tnoteref{label1}}
%% \tnotetext[label1]{}
%% \author{Name\corref{cor1}\fnref{label2}}
%% \ead{email address}
%% \ead[url]{home page}
%% \fntext[label2]{}
%% \cortext[cor1]{}
%% \address{Address\fnref{label3}}
%% \fntext[label3]{}

\title{Exact solution of long-range  electron transfer through conjugated molecular bridge.}

%% use optional labels to link authors explicitly to addresses:
%% \author[label1,label2]{}
%% \address[label1]{}
%% \address[label2]{}

\author{Ashish Kumar, Diwaker, Anirudhha Chakraborty }

\address{School of Basic Sciences, Indian Institute of Technology Mandi, Mandi, Himachal Pradesh 175001, India}

\begin{abstract}
Intermolecular electron transfer reaction often occurs over long range distances (i.e. up to several tens of angstroms) and plays a key role in various physical, chemical and biological processes. In these reactions the rate constant of long range electron transfer depends upon electronic coupling between the donor and acceptor. The coupling between donor and acceptor may increases by the atoms located between them which form a kind of bridge for electron tunneling. By using exact analytical method we calculated the value of electronic coupling for the above said processes in which the interaction of an electron with the donor, acceptor are  represented as Dirac delta functions and conjugated bridge is represented by finite square well.
\end{abstract}
\begin{keyword}
electron transfer rate , electronic coupling, distance between donor and acccetor.
%% keywords here, in the form: keyword \sep keyword
%% PACS codes here, in the form: \PACS code \sep code
%% MSC codes here, in the form: \MSC code \sep code
%% or \MSC[2008] code \sep code (2000 is the default)
\end{keyword}
\end{frontmatter}
%% \linenumbers
%% main text
\section{Introduction}
Intermolecular electron transfer (i.e either between different freely diffusing, donor and acceptor molecules or between Donor (D) and acceptor (A) subunits of supermolecule through bridging (B) subunit of supermolecule) which occurs over long distances ranging over several 10 of angstrom plays vital role in various scientific phenomenons which include physical chemical and biological processes. Intermolecular electron transfer (IT) is one of the  fundamental process of molecular electronics and applied research field. Natural Photosynthesis is the one of the best example, in which light harvesting molecules gather and transfer energy to reaction centers, where cascades of electron transfer reactions take place. Redox reactions and respiration are also some of the well cited examples in literature\citep{David, Moore, Satyam}. Such kind of long range electron transfer reactions are termed as Bridge mediated electron transfer (IT)reactions which involve the tunneling of an electron from a localized donor state to acceptor state by an intervening bridge that connects the donor and the acceptor. The internuclear distance between donor and acceptor and  electronic structure of bridge component in D-B-A system are well known to play a crucial role in determining the electron transfer rate\cite{McConnell, Barbara, Wasielewski, William, B}. The electron transfer rate has exponential behavior with the internuclear distance between donor and acceptor or length of bridge molecules in most of D-B-A systems studied by the many authors\cite{B, J, V.P.Zhdanov}. In biological processes the bridge is often a protein molecule and distance which electron had to transfer from donor to acceptor is usually greater than 6 angstrom. In intermolecular electron transfer processes the electron transfer rate explicitly depends on communication between donor and acceptor i.e. electronic coupling ($ G_{DA}$). The current work deals with the calculation of exact analytical formula for  the electronic coupling between donor and acceptor where the electronic coupling is treated as a function of internuclear distance R between the donor and the acceptor and length(L) of conjugated bridge. The donor and acceptor are both represented as Dirac delta Potentials  while the conjugated chain of atoms (i.e. bridge) which is responsible for electron transfer is represented as a finite square well. There are different experimental, analytical and computational approaches used by different authors to explain the electron transfer process through bridge between donor and acceptor in intermolecular electron transfer processes. For example in molecular electronics, Bo Albinsson and co-authors\cite{B} studied the electron and energy transfer mediated by the $\pi$-conjugated molecular bridges in D-B-A system and specifically studied the influence of donor, bridge structure and distance between donor and acceptor on the electronic coupling by using different bridge molecules. Yinxi Yu and co-workers have studied  bridge-mediated inter-valence electron transfer coupling in different metallocene complexes\cite{Y} by using computational approach. In his work he used constrained density functional theory to study the impact of the bridge length on the electronic coupling between the metal centers by using different models. Jean-Luc Brendas and Co-workers\cite{J} also studied the charge and energy transfer in $\pi$-conjugated oligomers and polymers. V.P. Zhdanov\cite{V.P.Zhdanov} in his finding used an analytical approach in which he derived an asymptotic expression for $G_{DA}$ for bridge-mediated electron process where he represents the bridge by single Dirac-delta function potential. Natalie Gorczak et.al\cite{N. G.} gives different mechanisms for electron and hole transfer along identical oligo-p-phenylene molecular bridges of increasing length. The bridge-mediated electron transfer processes studied by many other authors\cite{S.Larsson, J.Wolfgang, I.A, A.Onipko, S.C, A, D.J} and they all  devolved different approaches to calculate the electronic coupling($G_{DA}$) which is the key parameter of electron transfer rate in bridge-mediated electron transfer processes. In our model we have considered the electron transfer through conjugated bridge because conjugated bridge has an advantage over non-conjugated one\cite{Y}. Our model for long-range electron transfer through conjugated bridge has an advantages over earlier discussed models which we will represent in results section.
 Furthermore the rate of this non adiabatic electron transfer(IT) can be calculated for any system by Fermi Golden rule\cite{N.S, J.J, J.I, M.Bixon} The schematic diagram for our model is shown in Figure 1.
\begin{equation}
R_{IT}= 2\pi|G_{DA}|^2(FC)
\end{equation}
where $G_{DA}$ is the electron coupling between the donor and the acceptor, (FC) is franck Condon factor\cite{R.A} with $\hbar=1$, m=1
\section{Formulation of the Problem} 
We consider one dimension electron transfer from donor (located at $ x=-\eta$ \;($\eta\equiv R/2$) to acceptor (located at $x=\eta$ ($\eta\equiv R/2$) and conjugated bridge formed by particles centered at $x=0$. R is the distance between donor and acceptor. We assumed that electron weakly interact with bridge in comparison to donor and acceptor. The electronic potential energy is represented by the following equation
\begin{equation}
V(x) = -\alpha\delta(x+\eta)-\alpha\delta(x-\eta)+U(x)
\end{equation}
The first and second terms in the above equation represents the position of donor and acceptor as represented by Dirac delta potentials and third term represents bridge by the finite square well potential because we are dealing with the conjugated molecular bridge in our model and finite square well potential can be used to represent the conjugated chain of atoms. Furthermore
\begin{equation}
U(x) = \left\{\begin{array}{ll} 
-V_0   &;for -a<x<a \\ 
 0     &;$ elsewhere$
\end{array}\right.
\end{equation}
\\ where $V_0 $ is depth of the well and $\alpha$ is the strength of Dirac delta functions and $ L=2a $is the width of the square well. The time-independent Schr\"odinger equation for the above potential can be written as
\begin{equation}
-\frac{1}{2}\frac{d^2\psi}{dx^2}+V(x)\psi=E\psi
\end{equation}
The solution of the above equation for bound states can be written as
For region $x>\eta$,  V(x)= 0 so
\begin{equation}
\frac{d^2\psi}{dx^2}=-2E\psi=k^2\psi,
\end{equation}
where
\begin{equation}
k=\sqrt{-2E}
\end{equation}
the general solution of Eq.(5) can be written as 
\begin{equation}
\psi(x)=A\exp[-kx]+B\exp[kx]
\end{equation}
 for x$\rightarrow\infty$,B=0 hence
\begin{equation}
\psi(x)=A\exp[-kx]
\end{equation}
For region $a<x<\eta$,\;V(x)= 0 the solution of Eq. (5) is
\begin{equation}
\psi(x)=C\exp[-kx]+D\exp[kx]
\end{equation}
For region $-a<x<a$,\hspace{10pt} V(x) = U(x) the solution of Eq. (4) is
\begin{equation}
\psi(x)=I\cos(lx)+F\sin(lx)
\end{equation}\\ where l is wave vector and is given by  \begin{equation}
l = \sqrt{2(E+V_0)} 
\end{equation} 
For region $-\eta<x<-a$,\;V(x) = 0 hence the solution of Eq. (5) is 
\begin{equation}
\psi(x)=G\exp[-kx]+H\exp[kx]
\end{equation}
For region $x<-\eta $\;V(x) = 0. the solution of Eq. (5) is 
\begin{equation}
\psi(x)=J\exp[kx]
\end{equation} 
Now by using the two boundary conditions i.e. wave function is continuous every where and its derivative is also continuous at every where except at the points where potential is infinite and the fact that since used potential is a symmetric function, so we can assume with no loss of generality that the solution are either even or odd and apply the boundary conditions on either side ( say, at $x=\eta , x= a$) or to the other side ( say, at $x= -\eta , x= -a$). Firstly the solution are calculated for ( i.e symmetric wave function)in case of above potential and hence the wave function in collective form can be written as 
\begin{equation}
   \psi_{S}(x)\propto\left\{\begin{array}{ll}
     \exp[-kx]             &;x>\eta\\
     C\exp[-kx]+D\exp[kx]  &; a<x<\eta\\
     I\cos(lx)              &;-a<x<a\\
     C\exp[kx]+D\exp[-kx]  &;-\eta<x<-a\\
     \exp[kx]              &;x<-\eta
     \end{array}\right.
     \end{equation}
continuity of $\psi_s(x)$ and discontinuity of derivative of $\psi_s(x)$ at  $x=\eta$ , gives
\begin{equation}
\exp[- k\eta]= C \exp[-k \eta]+D \exp[K \eta]
\end{equation}
and
\begin{equation}
-k\exp[-k\eta]-[-C k\exp[-k \eta]+D k\exp[k \eta]= -2\alpha\exp[-k \eta]
\end{equation}
further simplification of these equations will give me 
\begin{equation}
1-C-D\exp[2 k \eta]=0
\end{equation}
and
\begin{equation}
\left(1-\frac{2\alpha}{k}\right)-C+D\exp[2k\eta]=0
\end{equation}
similarly at points x=a, another set of equations can be written as
\begin{equation}
C\exp[-ka]+D\exp[ka]=I\cos(la)
\end{equation}
and
\begin{equation}
-kC\exp[-ka]+kD\exp[ka]=-lI\sin(la)
\end{equation}
Then from Eq. (17) and Eq. (18) we obtained.
\begin{equation}
D=\frac{\alpha}{k}\exp[-2k\eta]
\end{equation}
and
\begin{equation}
C=\left(\frac{k-\alpha}{k}\right)
\end{equation}
 and by Dividing Eq. (20) by Eq. (19), we get
 \begin{equation}
 \tan(la)=\frac{k}{l}\left(\frac{C\exp[-ka]-D\exp[ka]}{C\exp[-ka] + D\exp[ka]}\right)
 \end{equation}
 Now by using Eq. (22) and Eq. (21) in above equation we get the analytical formula
\begin{equation}
\tan(la)=\frac{k}{l}\left(1-\frac{2\alpha\exp[2k(a-\eta)]}{k-\alpha+\alpha\exp[2k(a-\eta)]}\right)
\end{equation}
This is the formula for allowed energies, since k and $l$ are both function of Energy. To simplify above expression we make the following substitution for our convenience: Let\\
 \begin{equation} z_s=la \;and \;z_0=a\sqrt{2V_0},\;
(k^2+l^2)=2V_0, and\; ka=\sqrt{z_0^2-z_s^2}
\end{equation} 
Using these substitutions Eq. (24) can be written as
 $$
\tan(z_s)=\sqrt{\left(\frac{z_0}{z_s}\right)^2-1}\left(1-\frac{2\alpha a\exp{[2(\sqrt{z_0^2-z_s^2}-k\eta)}]}{\sqrt{z_0^2-z_s^2}-\alpha a+\alpha a\exp{[2(\sqrt{z_0^2-z_s^2}-k\eta)}]}\right)
$$
further 
\begin{equation}
\tan(z_s)=\sqrt{\left(\frac{z_0}{z_s}\right)^2-1}\left(1-\frac{\alpha L \exp[2\sqrt{z_0^2-z_s^2}(1-\frac{R}{L})]}{\sqrt{z_0^2-z_s^2}-\frac{\alpha L}{2}+\frac{\alpha L}{2}\exp[2\sqrt{z_0^2-z_s^2}(1-\frac{R}{L})]}\right)
\end{equation} Here $L=2a$ is the length of conjugated chain of atoms between donor and acceptor (i.e bridge).\\
This is the transcendental equation for $z_s$\;(and hence for $E_s$) as function of $z_0$ (which is a measure of the size of the well). The above equation is solved graphically to get the value of $z_s$ and hence of energy(E) for  symmetric states.\\
Similarly we can adopt the same methodology for antisymmetric states as mentioned below. The antisymmetric wave function is represented as
\begin{equation}
  \psi_{u}(x)\propto\left\{\begin{array}{ll}
     \exp[-kx]              &;x>\eta\\
     C\exp[-kx]+D\exp[kx]   &; a<x<\eta\\
     F\sin(lx)              &;-a<x<a\\
    -C\exp[kx]-D\exp[-kx]   &;-\eta<x<-a\\
    -\exp[kx]               &;x<-\eta
     \end{array}\right.
 \end{equation}
 applying the boundary conditions at $x=\eta$ and $x=a$ we get corresponding transcendental equation  for $z_u$ (and hence for $E_u$ ).
   \begin{equation}
  -cot(z_u)=\sqrt{\left(\frac{z_0}{z_u}\right)^2-1}\left(1-\frac{\alpha L \exp[2\sqrt{z_0^2-z_u^2}(1-\frac{R}{L})]}{\sqrt{z_0^2-z_u^2}-\frac{\alpha L}{2}+\frac{\alpha L}{2}\exp[2\sqrt{z_0^2-z_u^2}(1-\frac{R}{L})]}\right)
 \end{equation}
  where \\ \begin{equation}
 z_u=la \hspace{10pt} \& \hspace{10pt}  z_0=a\sqrt{2V_0}
 \end{equation}
 This is another transcendental equation for $z_u$(and hence for $E_u$) as function of $z_0$ which is solved by graphical method to get the value of energy(E) for antisymmetric states.\\ The electronic coupling between the donor and the acceptor equals to half the energy differences of lowest antisymmetric and symmetric states\cite{V.P.Zhdanov}.i.e
 \begin{equation}
 G_{DA} =\frac{\left(E_u-E_s\right)}{2}
 \end{equation}
 using the values from Eq. (11), (25) and (29) we get,
\begin{equation}
E_u=\frac{z_u^2}{2a^2}-V_0\;and\;E_s=\frac{z_s^2}{2a^2}-V_0 
\end{equation}
hence Eq. (30) becomes:\\
 \begin{equation}
 G_{DA} =\frac{\left(z_u^2-z_s^2\right)}{L^2} 
 \end{equation}
This is expression for electronic coupling for system having conjugated bridge. Now we analyses the case where energy of the bridge level is appreciably higher than the energies of lowest antisymmetric and symmetric states.
\section{Results}
  We choose the parameter  $ \alpha = \frac{1}{\sqrt{2}}, LV_0 = \frac{2}{\sqrt{3}}$ such that energy of donor, acceptor and bridge is comparable to the binding energies in real electron transfer reactions\cite{V.P.Zhdanov} and analyze our formalism for electronic coupling vs internuclear distances between donor and acceptor ranging from (16 - 50) atomic units. The results for electronic coupling are shown in Figure (2a, 2b, 3a, 3c). Figure 2a shows the exponential variation of electronic coupling vs internuclear distances between donor and acceptor calculated by us for conjugated  bridge mediated electron transfer processes while solving Eq. (26) and Eq. (28) by graphical method and plugging the results into Eq. (30). The reason for choosing this range is that electron transfer reactions in proteins and other conjugated systems generally occurs within this range\cite{David}. Figure 2b shows the logarithmic variation of electronic coupling vs internuclear distances between donor and acceptor. The reason for taking log values is to compare our exact analytical results with the already published results for such type of coupling by others i.e Figure 3a\cite{V.P.Zhdanov}. From this comparison we conclude that the value of electronic coupling calculated by other authors comes in the range from $10^{-4}\; to \;10^{-14}$ atomic units by varying the internuclear distances from (16-50) atomic units, however using our exact analytical results this values ranges from $10^{-1}\; to \;10^{-2}$ atomic units over same range of internuclear distances between donor and acceptor thereby showing that the value of electronic coupling is better in our case. Figure 3b shows the area of direct overlapping( i.e. the area of overlapping between donor and acceptor wave functions shown in Figure 4) in atomic units vs internuclear distances between donor and acceptor for the bridge mediated electron transfer processes which shows that if we move below the 16 atomic units then the electronic transfer may occurs by direct overlapping instead of bridge between the donor and the acceptor. The figure 3c shows that how the electronic coupling ( which is the exponential function of internuclear distances(R) between donor and acceptor) is  decreasing with increasing length of different  conjugated bridges having different length. Our results which shown in figure 2e are qualitatively similar to the earlier published experimental and computational results in Fig.14\cite{B} for different conjugated bridges ( i.e. Oligoethynylenes (OE), Phenyle-Oligoethynylenes (Ph-OE) and Oligo-p-phenyleneethynylene (OPE)) having different length for same donor and acceptor(Pentacene) bridge-mediated electron transfer systems. In both cases  value of electronic coupling ($G_{DA}$)  decreases as length of bridge increases. Our results have same behavior as the earlier published computational results shown in Table 6.\cite{Y} by Yinxi Yu and coworkers, which shows  that the value of electronic coupling  decreases with increases the length of bridge for bridge-mediated intervalence electron transfer in different metallocene complexes. Our model for bridge-mediated electron transfer processes has advantages over earlier model because our model is universal model and it can be use to calculate or estimate the value of electronic coupling ($ G_{DA}$) without doing any computational or experimental work for any conjugate bridge-mediated electron transfer processes if we know the values of binding energies of donor/acceptor molecule and bridge molecule of the systems( e.g. supermolecules) involve electron transfer processes. Hence we can calculate the rate of electron transfer for bridge-mediated electron transfer processes for such systems.
 \section{Conclusion}
 On comparison of results obtained by our exact analytical method in which we assumed that electron transfer through conjugated bridge in electron transfer processes with the already published results for such type of coupling by author\cite{V.P.Zhdanov} we conclude results obtained by our method are better and
qualitatively similarity of our results with the earlier published experimental and computational results by authors\cite{B,Y} we conclude that our model is easier model to calculate the value of electronic coupling($G_{DA}$) hence rate of electron transfer for conjugated bridge-mediated electron transfer processes. Thus our exact analytical method  is better to calculate the  long-range conjugated bridge-mediated electron transfer rate for real physical, chemical and biological processes. 
\section*{Acknowledgements}
 The authors would like to thank  School of Basic Sciences, Indian Institute of Technology Mandi, Mandi, Himachal Pradesh 175001, India for financial support and providing necessary facilities to accomplish this work.
%% The Appendices part is started with the command \appendix;
%% appendix sections are then done as normal sections
%% \appendix
%% \section{}
%% \label{}
%% If you have bibdatabase file and want bibtex to generate the
%% bibitems, please use
%%
%%  \bibliographystyle{elsarticle-harv} 
%%  \bibliography{<your bibdatabase>}
%% else use the following coding to input the bibitems directly in the
%% TeX file.

\begin{figure}
\centering
\includegraphics[width=10cm,height=8cm]{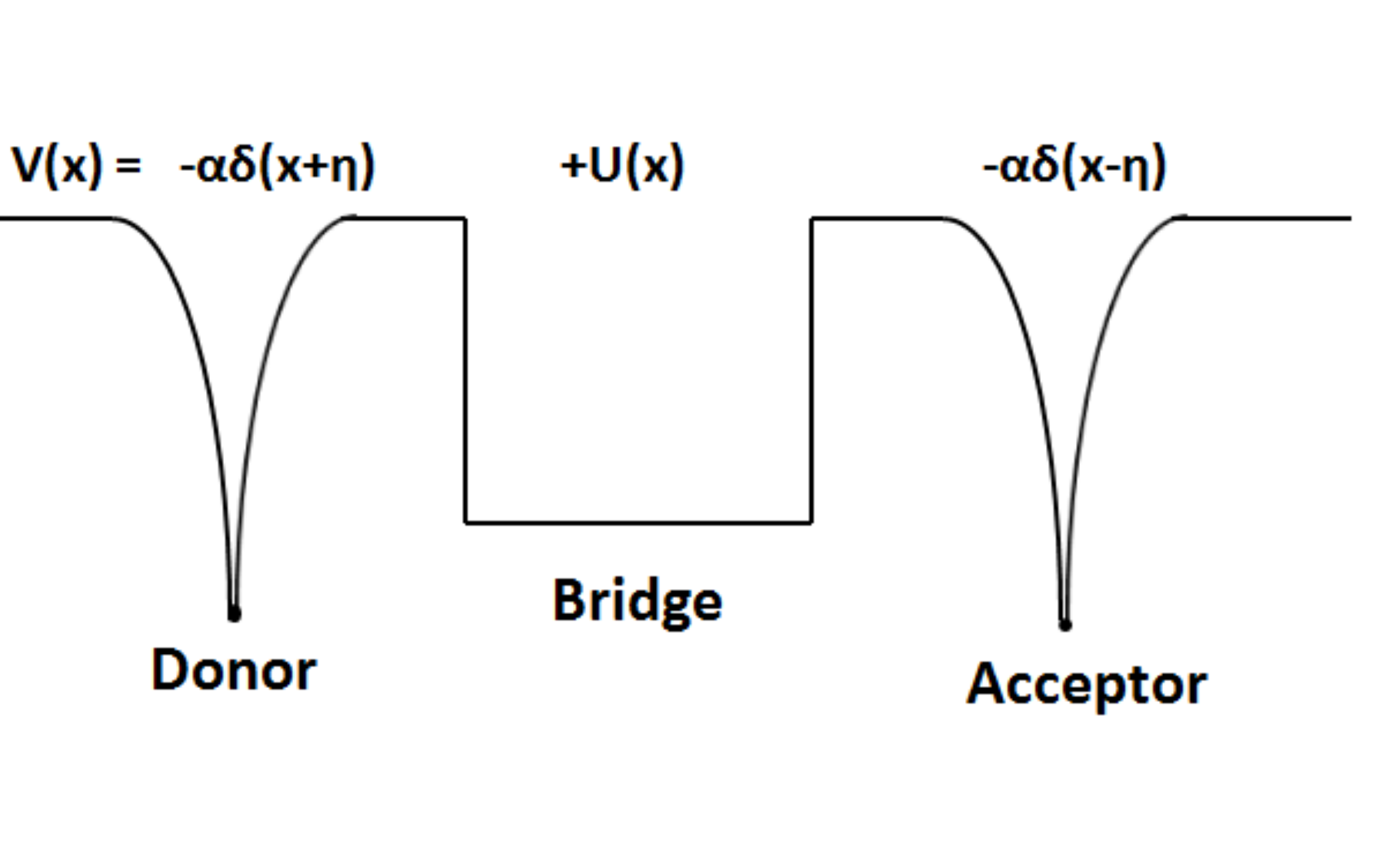}
\caption{Schemaic diagram for our model}
\end{figure}
\begin{figure}
 \subfigure[Electronic coupling($G_{DA}$) between donor and acceptor  varies as exponential  function of atomic distance(R)  between them for conjugated bridge-mediated electron transfer processes]{\includegraphics[width=80mm]{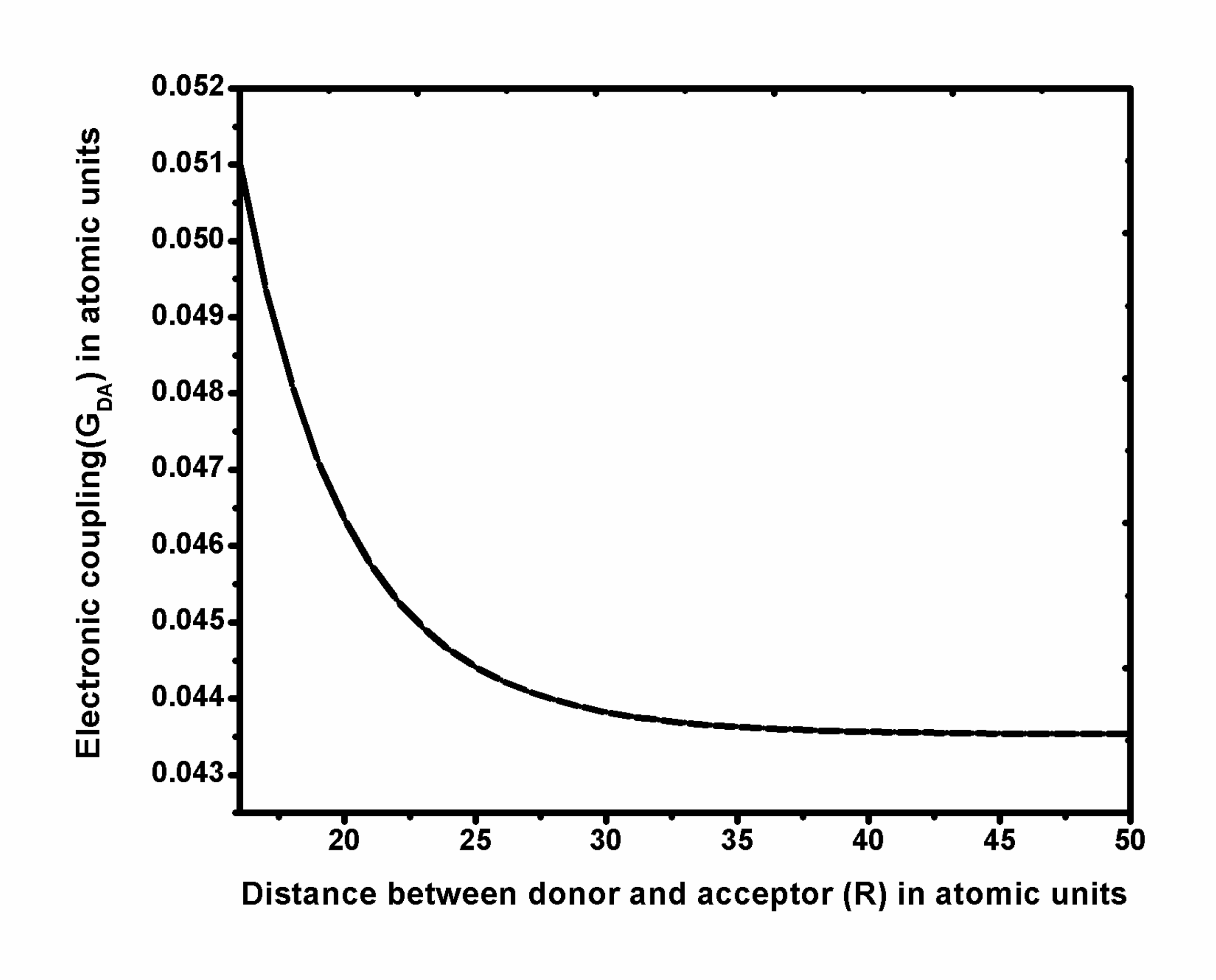}\label{first}}
 \subfigure[Electronic coupling($G_{DA}$) as exponential function of  distance between donor and acceptor (R)  for electron transfer through conjugated bridge in IT processes]{\includegraphics[width=80mm]{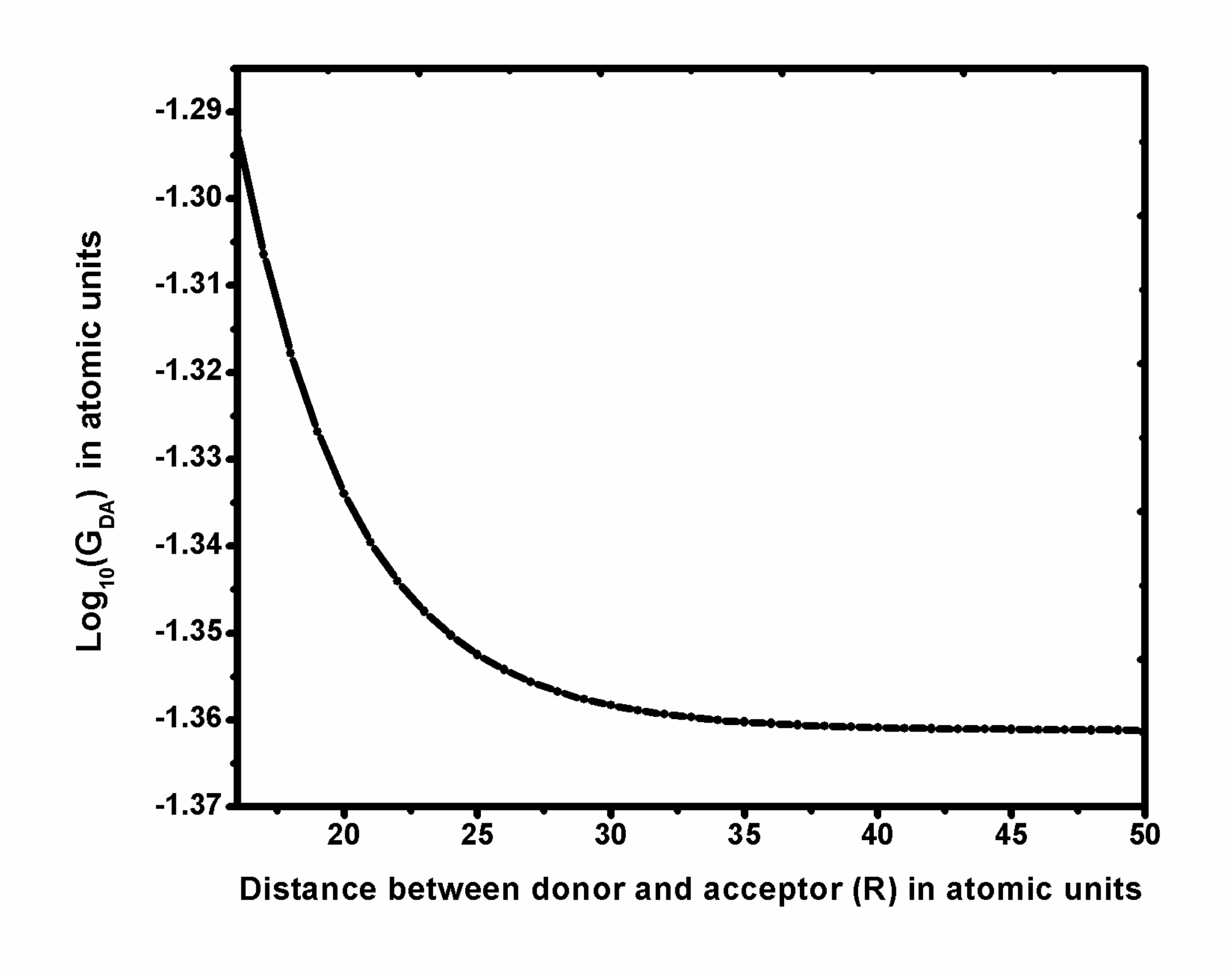}\label{second}}
 \caption{Electronic coupling vs Internuclear distance}\label{main_label}
 \end{figure}
 \begin{figure}
 \subfigure[Electronic coupling ($G_{DA}$) as exponential function of  distance between donor and acceptor 
(R) for electron transfer through bridge represented by single Dirac-delta function (i.e. non-conjugated bridge) in IT processes]{\includegraphics[width=80mm]{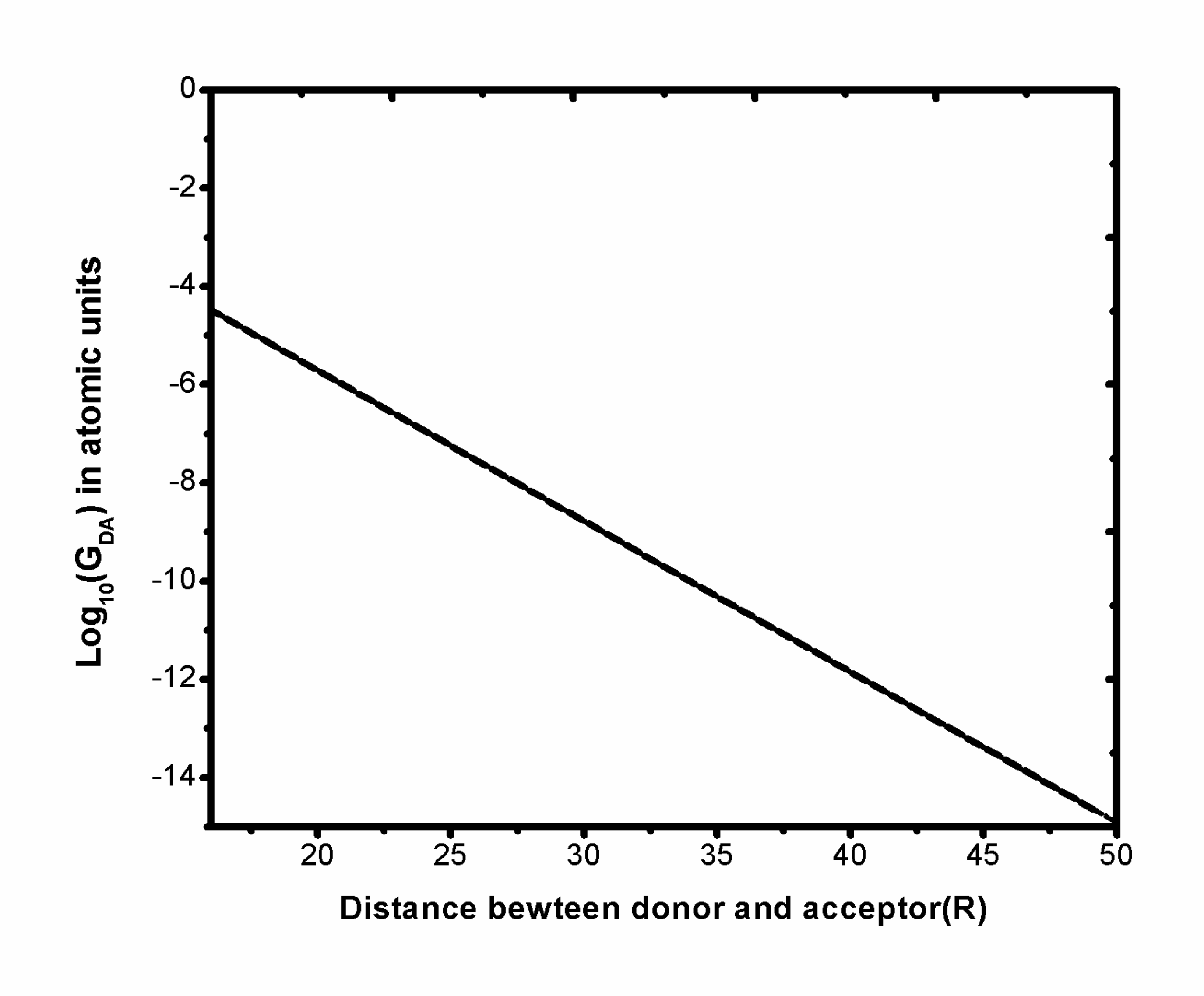}\label{first}}
 \subfigure[Area of direct overlapping between donor and acceptor as function of atomic distances between donor and acceptor from R=2a.u to 15 a.u.]{\includegraphics[width=80mm]{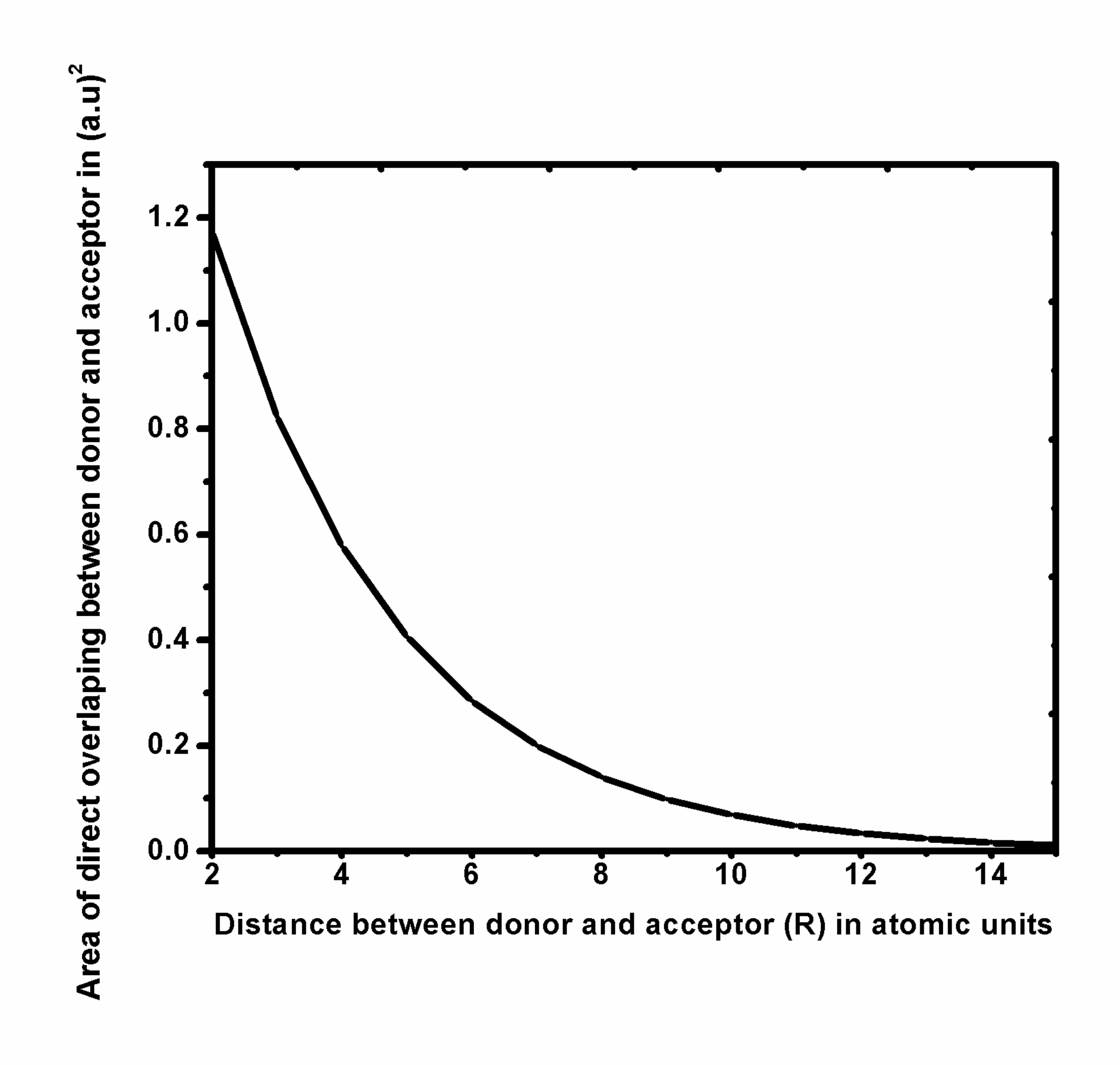}\label{second}}
 \begin{center}
  \subfigure[Electronic coupling($G_{DA}$) between donor and acceptor varies as exponential function of atomic distances(R) between them for different conjugated bridge having different lengths($L1 < L2 < L3 < L4$)]{\includegraphics[width=100mm]{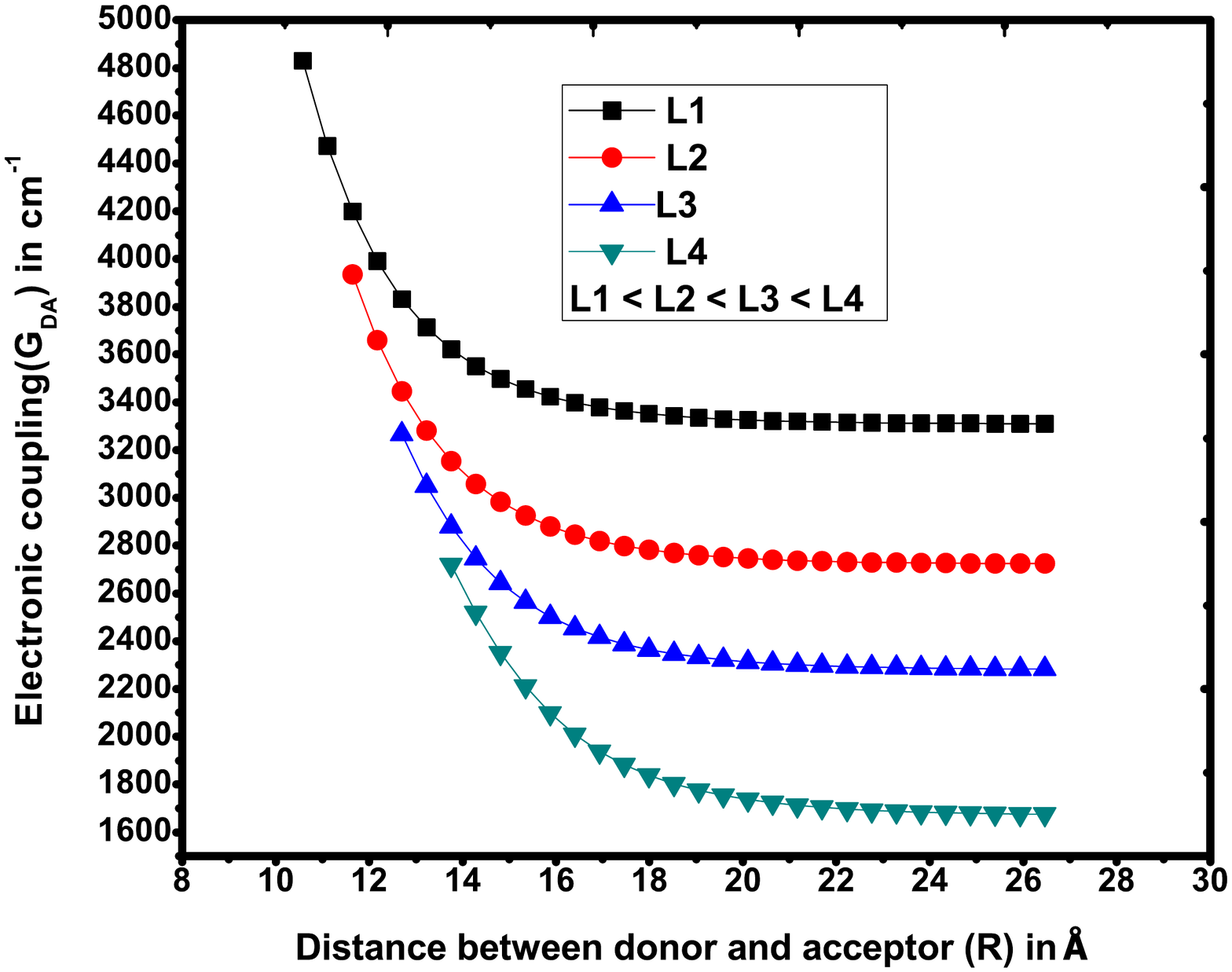}\label{first}} 
 \end{center}
 \caption{Electronic coupling vs Internuclear distance}\label{main_label}
\end{figure}
 \begin{figure}
\centering
\includegraphics[width=10cm,height=8cm]{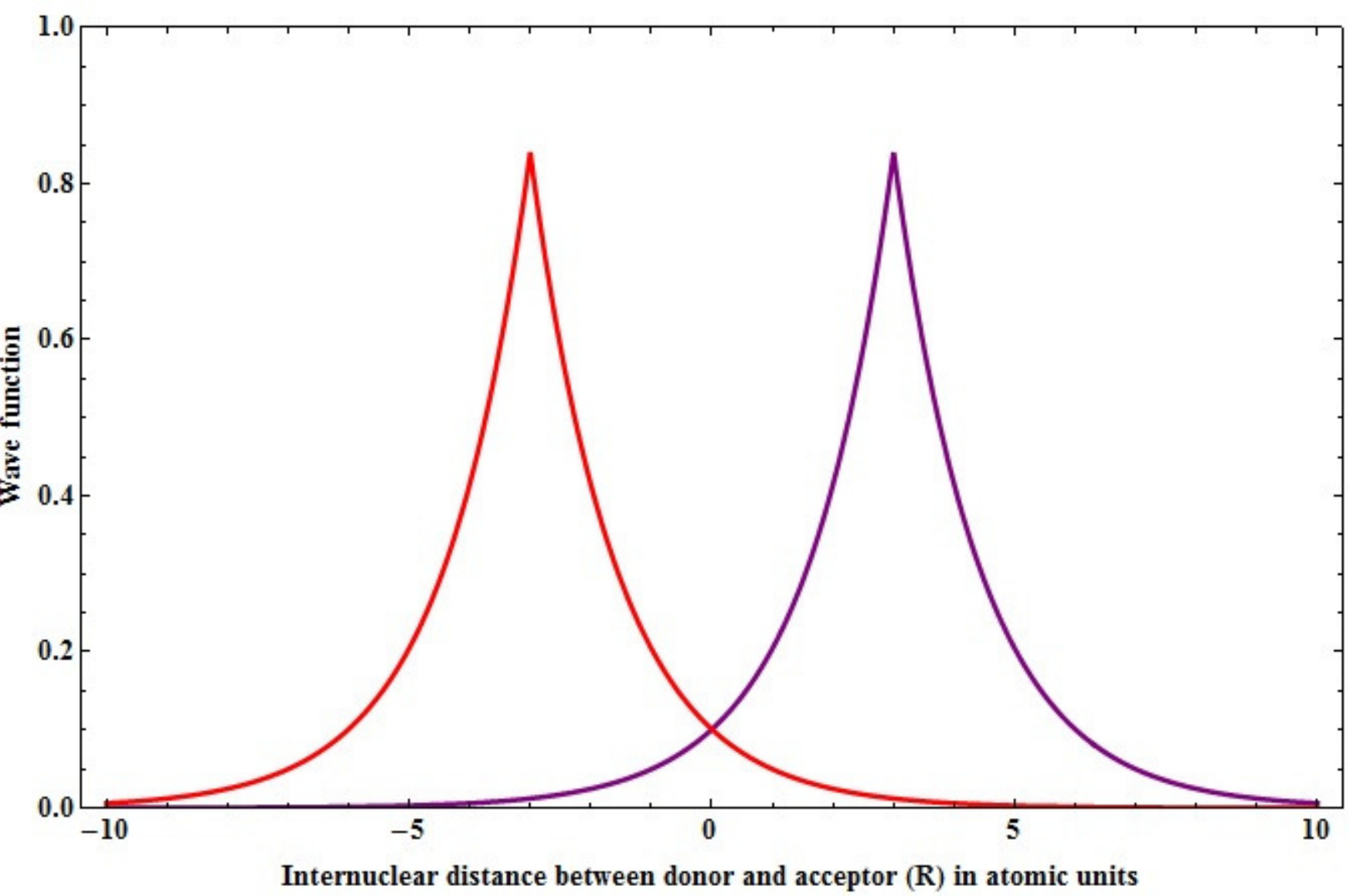}
\caption{Overlapping between donor and acceptor wave functions having internuclear distance 8 a.u}
\end{figure}
\end{document}